# SIMULATION OF PARTICLE SIZE EFFECT ON DYNAMIC PROPERTIES AND FRACTURE OF PTFE-W-Al COMPOSITES

E.B. Herbold[1], J. Cai[2], D. J. Benson[1] and V.F. Nesterenko[1,2]

[1] *Dept. of Mechanical and Aerospace Engineering, University of California, San Diego, La Jolla, CA 92093-0411*
[2] *Materials Science and Engineering Program, University of California, San Diego, La Jolla, CA 92093-0418*

**Abstract.** Recent investigations of the dynamic compressive strength of cold isostatically pressed composites of polytetrafluoroethylene (PTFE), tungsten (W) and aluminum (Al) powders show significant differences depending on the size of metallic particles. The addition of W increases the density and overall strength of the sample. To investigate relatively large deformations multi-material Eulerian and arbitrary Lagrangian-Eulerian methods, which have the ability to efficiently handle the formation of free surfaces, were used. The calculations indicate that the increased strength of the sample with fine metallic particles is due to the formation of force chains under dynamic loading. This phenomenon occurs even at larger porosity of the PTFE matrix in comparison with samples with larger particle size of W and higher density of the PTFE matrix.

Tailoring the mechanical and chemical properties of reactive materials is important and necessary for various applications. For example, this can be accomplished by varying particle size and morphology in plastic bonded explosives (PBX) [1-6] or layer thicknesses in laminate or flake composites [2,7,8]. Recent literature [2, 3] investigates highly energetic reactions between PTFE and Aluminum. It is important that the reaction in the bulk of this material is driven by mechanical deformation.

The following numerical investigation focuses on mechanical properties of a reactive composite processed using cold isostatic compression (CIP) of a Teflon (PTFE), Aluminum (Al) and Tungsten (W) mixture where the PTFE acts as a soft binder. PTFE-W-Al mixtures were investigated in quasi-static, Hopkinson bar, and Drop-weight experiments [9, 10]. The particle size and distribution within the sample were varied to investigate the resulting failure mechanisms and sample strength. The sample density ranged from 6 g/cm$^3$ to 7.2 g/cm$^3$.

In drop weight tests [10] the average compressive strength of the porous samples,

containing small W particles (< 1 μm) was 55 MPa compared to the denser samples (32 MPa), containing large W particles (< 44 μm). This behavior was also observed in quasi-static and Hopkinson bar tests [9]. These results may seem counterintuitive. A possible explanation is that force chains provide additional strength by creating an underlying material mesostructure. Force chains in granular packings of photoelastic discs as analogs of energetic materials were investigated in [4-6]. The percolation limit [11] of the constitutive grains within the matrix must be considered to determine the coordination number that is high enough to easily facilitate the formation of force chains or 'stress bridges'. These force chains may break and reform as the sample mesostructure self-organizes in response to the load.

## MODEL DESCRIPTION

In the following numerical investigation two samples of a PTFE-W-Al composite with different assemblies of metallic particles with the same volume fraction were used to investigate the effect of force chains, created by metallic particles, on the global mechanical response of the samples. The drop-weight test is suitable for comparison with results of numerical analysis of highly heterogeneous samples because of inherently simple boundary conditions: a constant velocity of the top of the sample.

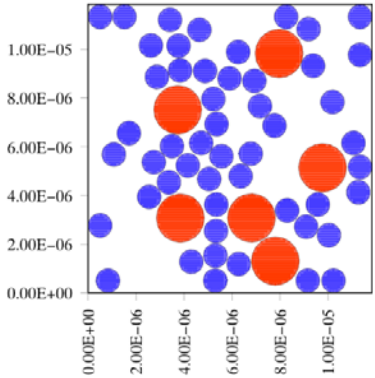

Figure 1. PTFE-W-Al sample using 2μm diameter Al particles and 1μm diameter W particles.

The two dimensional mesostructure in numerical analysis is used to estimate the level of influence that the force chains have on the global behavior of samples though it does not reproduce the coordination number of three dimensional packing at the same volume content of constituents. Two samples using a randomly distributed mixture of small (1 μm) W and Al particles (2 μm) (Fig. 1) and large (10 μm) W and Al particles (2 μm) (Fig. 2) are used in finite element calculations to investigate the force chain effect. These two different particle arrangements result in self-assembled force chains in former case and do not appear in the other one. The weight and volume fractions of each sample constituents were the same in both calculations (e.g. volume fractions: 59% PTFE, 28% W, 13% Al) and were close to the values in experiments [9].

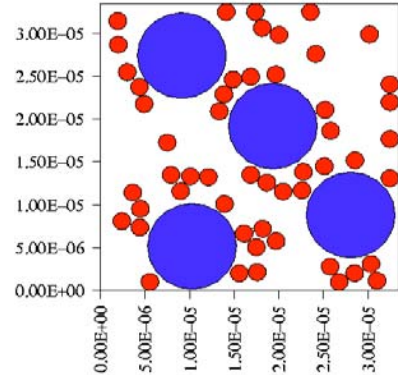

Figure 2. PTFE-W-Al sample using 2μm diameter Aluminum particles and 10μm diameter Tungsten particles.

A two-dimensional multi-material Eulerian Hydrocode [12] with diffusive heat transfer is implemented to investigate the quasi-static behavior of the sample. Each material in the mixture has different equations of state, physical and mechanical properties. The material model used for PTFE was the Johnson-Cook with Failure,

$$\sigma_y = \left[A + B(\bar{\varepsilon}^p)^n\right]\left[1 + C\ln\dot{\varepsilon}^*\right]\left[1 - T^{*m}\right] \qquad (1)$$

where $A$=11 MPa, $B$ = 44 MPa, $n$ = 1, $C$ = 0.12 and $m$ = 1 (extrapolated from the data given in [13]). The material failure strain was specified in a separate failure criterion where the failure strain was specified as $\varepsilon_f$ = 0.05. This value was obtained from quasi-static and Hopkinson bar experimental data of pure CIPed PTFE samples [9]. The Gruneisen equation of state was used to define the pressure in compression and tension.

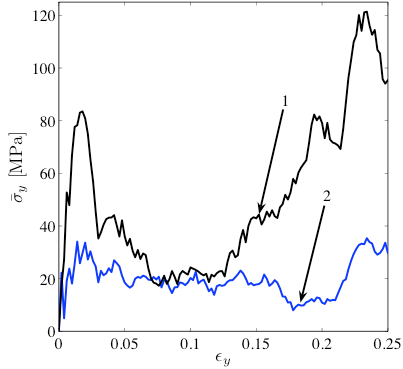

Figure 3. Average stress at the top of the sample plotted against the 'global' strain for a sample using small tungsten particles (curve 1) and a sample using large tungsten particles (curve 2). Note the stress increases in curve 1 after 0.13 global strain while the curve 2 coincides with the results for pure CIPed PTFE (not shown).

The Johnson-Cook material model without failure was used for the tungsten and aluminum particles. The Johnson-Cook parameters for tungsten are $A = 1.51$ GPa, $B = 177$ MPa, $n = 0.12$, $C = 0.016$, and $m = 1$. The Johnson-Cook parameters for aluminum are $A = 265$ MPa, $B = 426$ MPa, $n = 0.34$, $C = 0.015$, and $m = 1$. Since the particle deformation or fracture during dynamic loading is negligible in the mixture with PTFE we used linear elasticity for W and Al.

The first random sample (Fig. 1) was selected due to the appearance of a particle distribution initially banded such that 'through-thickness' force chains would immediately resist the loading. The large tungsten (10 μm diameter) and small aluminum (2 μm diameter) particles in the second random sample (Fig. 2) were distributed evenly throughout the PTFE matrix making the formation of force chains less probable reflecting the experimental difference in their tapped densities (discussed later).

In both samples, a constant velocity of 4.43m/s was prescribed along the top boundary and the bottom was fixed. The sample material was allowed to slip along both the top and bottom boundaries without friction. A tungsten slab was placed along the top and bottom boundaries to simulate the drop-weight test conditions where a steel anvil was used to strike the sample on a steel base. The sides of the samples were free from restriction.

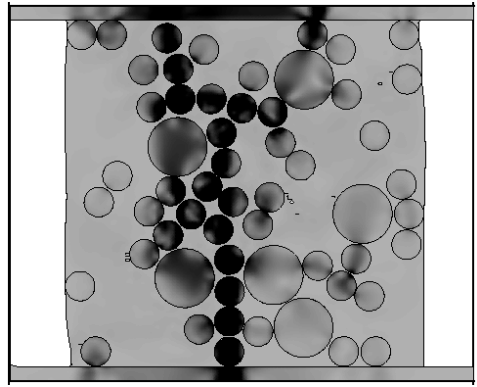

Figure 4. The von Mises stress distribution at 0.022 'global' strain. The stress intensity varies from 0 (light gray) to 50 MPa (dark gray).

The stress-strain plot for the two drop-weight calculations is shown in Fig. 3. The stress values in the vertical direction correspond to the average stress across the top boundary. The strain corresponds to the 'global' strain or difference in the height of the sample divided by the initial height. Notice that the effective 'first yield' of the porous sample with force chains (curve 1) is 85 MPa and about 38 MPa for the sample without force chains (curve 2).

The von Mises stress distribution (Figs. 4 and 5) and the global effective plastic strain (Fig. 6) for the sample with fine tungsten particles are shown. Figure 4 shows the stress distribution within the sample at 0.022 'global' strain. A single force chain is apparent starting from the top left-center through the bottom of the sample. This can be compared to the sudden increase in the stress strain plot shown in curve 1 in Fig. 3 at the corresponding strain. Upon further deformation, this force chain disintegrates resulting in the decrease in stress in curve 1 (Fig. 3). Force chains are reactivated (Fig. 5) upon further deformation when the global strain is 0.238. There are now two force chains; one beginning from the top left-center to the bottom and another force chain beginning just to the right of the center.

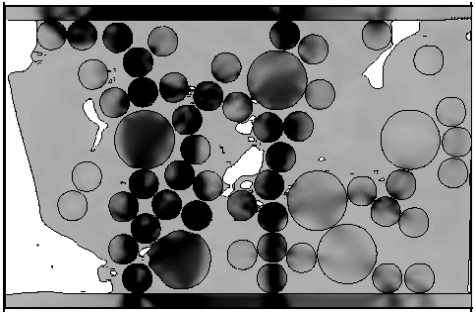

Figure 5. The von Mises stress distribution at 0.238 'global' strain for the sample with force chains. The stress intensity varies from 0 (light gray) to 50 MPa (dark gray).

This self organization of metallic particles was accompanied by a macrocrack formed diagonally from the top right to the bottom left at 0.238 'global' strain (Figs. 5 and 6). The local effective plastic strain in the sample (Fig. 6) above this crack shows that the damage in the PTFE matrix is distributed around the metal particles. It is important to maintain the damage throughout the sample bulk to enhance a possible chemical reaction.

The following features can be observed from the results of the calculation using the first sample that facilitated the presence of force chains. The global motion of the metallic particles is comparable to their sizes and the size of the samples. Such large displacements resulted in force chains being created, destroyed and reactivated (with different particles) in the course of sample deformation and fracture. It is interesting that the progressive fracture corresponds to the spikes of global stress (curve 1, Fig. 3) and the maximum global stress is observed for high strain levels in the heavily fractured sample. Metal particles initiated a shear macrocrack in the PTFE matrix propagating at 45 degrees to the loading direction. After the initiation of the macrocrack, part of the sample was left undeformed. This type of behavior should be avoided since the initiation of a reaction will not occur in such areas.

The second sample (Fig. 2) did not have a particle distribution conducive to force chain activation. 'Through-thickness' force chains are not present in the sample up to 0.25 global strain, though groups of particles have created localized chains. It is apparent from calculations (not shown) that macrocracks formed in the sample without force chains prohibited any bulk distributed damage.

Separate calculations (not shown) with a pure PTFE sample have a stress strain behavior very similar to curve 2 in Fig. 3, which suggests that the load was resisted by the matrix material alone in the second sample.

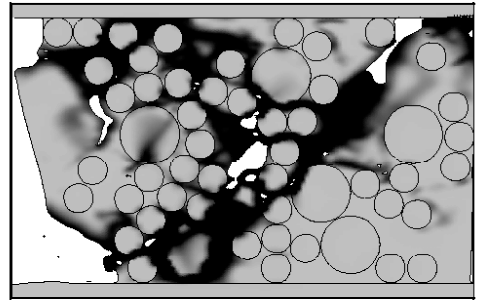

Figure 6. The effective plastic strain distribution at 0.238 'global' strain for the sample with force chains. The level of strain varies from 0 (light gray) to 0.05 (dark gray).

The global motions of the metal particles in the second sample are also comparable to their sizes and to the size of the sample. However such large displacements did not result in force chains. The stress spikes were not related to the activation of force chains propagating from top to bottom. The maximum stresses in the spikes are less than two times lower than in the previous sample. The metal particles initiated shear macrocracks in the PTFE matrix propagating at 45 degrees from the direction of compression. The bottom part of the sample remained mostly undeformed after the initiation of the macrocracks as it was also the case for the first sample (Fig. 4), which hinders reaction initiation in such areas.

The presented two-dimensional calculations demonstrated that force chains created by metallic particles are a probable cause of the higher strength of these mixtures. This is also supported by the comparison of the volume content of metal particles in mixtures with PTFE and the tapped densities of mixtures of coarse W and fine Al and fine W and fine Al powders [9]. For example, the volume content of fine W and fine Al particles in mixture with PTFE for a CIPed sample with density 6 g/cm$^3$ is 0.36. This is higher than volume content of solid (0.27) in the tapped mixture of fine W and fine Al taken at their mass ratio as in the mixture with PTFE. This means that the force chains supporting the mesostructure in this tapped powder will be also present in the CIPed composite sample.

This is not the case is with a CIPed composite sample using coarse W and fine Al (density 7.05 g/cm$^3$). Here the volume content of metal particles is significantly smaller (0.425) than volume content of solid (0.69) in the mixture of coarse W and fine Al powders at tapped density taken at the same mass ratio as in the mixture with PTFE. This means that PTFE matrix is dispersing metal particles preventing them from forming force chains in this sample.

These results show that it is less probable that the higher porous sample strength is due to the bulk distributed fracture in mixtures with fine W (Al) particles causing large interfacial area between metal particles and the PTFE matrix delaying the propagation of macrocracks and effectively increasing the global strength.

The results of two-dimensional numerical calculations correctly predict development of macrocracks in the composite sample at about 2% strain, which is within the range observed in experiments. It should be emphasized that this level of strain is smaller than failure strain in the PTFE matrix due to strain concentrations caused by rigid metal particles.

## ACKNOWLEDGEMENTS

The support for this project provided by ONR (Program Officer Dr. Judah M. Goldwasser, N00014-06-1-0263 and MURI ONR Award N00014-07-1-0740) is highly appreciated.